\documentclass[journal=jctcce,manuscript=article]{achemso}
\pdfoutput=1
\usepackage[version=3]{mhchem} 
\usepackage{xcolor}
\usepackage{amssymb,amsmath,amsfonts}
\usepackage{mathtools}
\usepackage{siunitx}

\newcommand{\pr}[1]{\mathbb{P}\left\{#1\right\}}

\renewcommand{\l}{\left}
\renewcommand{\r}{\right}

\newcommand{\red}[1]{{\color{red}{#1}}}

\newcommand{\YT}[1]{{\red{\bf YT: #1 }}}
\newcommand{\dd}{\text{d}}
\definecolor{Purpleee}{RGB}{91,40,224}

\newcommand{\comment}[1]{}

\author{Van A. Ngo}
\email{ngoav@ornl.gov}
\affiliation[ORNL]{Advanced Computing for Life Sciences and Engineering, Computing and Computational Sciences, National Center for Computational Sciences, Oak Ridge National Lab, Oak Ridge, TN 37830}
\affiliation[CNLS]{Center for Nonlinear Studies (T-CNLS), Theoretical Division, Los Alamos National Laboratory, NM 87545, USA}
\author{Yen Ting Lin}
\affiliation[CCS]{Information Sciences Group (CCS-3), Computer, Computational and Statistical Sciences Division, Los Alamos National Laboratory, NM 87545, USA}
\email{yentingl@lanl.gov}
\author{Danny Perez}
\affiliation[T]{Physics and Chemistry of Materials Group (T-1), Theoretical Division, Los Alamos National Laboratory, NM 87545, USA}
\email{danny_perez@lanl.gov}

\title[Koopman operator with Kolmogorov-Smirnov indicator functions ]
  {Improving Estimation of the Koopman Operator with Kolmogorov-Smirnov Indicator Functions}

\begin{document}


\begin{abstract}

It has become common to perform kinetic analysis using approximate Koopman operators that transforms high-dimensional time series of observables into ranked dynamical modes. Key to a practical success of the approach is the identification of a set of observables which form a good basis in which to expand the slow relaxation modes. Good observables are, however, difficult to identify {\em a priori} and sub-optimal choices can lead to significant underestimations of characteristic timescales. Leveraging the representation of slow dynamics in terms of Hidden Markov Model (HMM), we propose a simple and computationally efficient clustering procedure to infer surrogate observables that form a good basis for slow modes. We apply the approach to an analytically solvable model system, as well as on three protein systems of different complexities. We consistently demonstrate that the inferred indicator functions can significantly improve the estimation of the leading eigenvalues of the Koopman operators and correctly identify key states and transition timescales of stochastic systems, even when good observables are not known {\em a priori}.

\end{abstract}

\section{Introduction}

Elucidating the kinetics describing rare structural or chemical reactions is crucial to understand many biophysical and biochemical systems \cite{noe2008TransitionNetworksModeling,pande2010EverythingYouWanted,prinz2011MarkovModelsMolecular,chodera2014MarkovStateModels,wu2016MultiensembleMarkovModels}. Even when long fully-resolved trajectories are available, e.g., via extensive molecular dynamics (MD) simulations, extracting a reliable representation of kinetics in terms of a handful of physical observables can be elusive due to the high dimensionality and complexity of most application-relevant systems, particularly in the absence of an intuitive reaction coordinate. A powerful approach aimed at tackling such problems exploits the correspondence between the spectral properties of the so-called Koopman operator and those of the dynamics generator. It allows to efficiently reduce high dimensional timeseries into a compact and tractable representation (See Sec.\ \ref{sec:Theory}). 
Various techniques and spectral analysis methods based on this mathematical formalism include time-independent component analysis (TICA) for improving Markov State Modeling (MSM)\cite{molgedey1994SeparationMixtureIndependenta,schwantes2013ImprovementsMarkovState,perez-hernandez2013IdentificationSlowMoleculara,naritomi2011SlowDynamicsProtein}, the variational approach for Markov processes (VAMP) incorporated in MSMbuilder\cite{beauchamp2011MSMBuilder2ModelingConformational,pandeMSMbuilder3} and PyEMMA\cite{scherer2015PyEMMASoftwarePackage}, as well as different variants of the extended dynamical mode decomposition (EDMD) approach \cite{williams2015data}. These methods underscore that analyzing the spectral properties of the Koopman operator is powerful to understand and characterize the dynamics of complex systems.

In essence, an estimated Koopman operator based on EDMD (or VAMP or TICA) is an approximation of the dynamics generator of measurable observables in an infinite-dimensional Hilbert space\cite{Koopman315,Koopman255,mezic2005a}. This approximate Koopman operator returns the expectation value at a time $t+\tau$ from a value of an observable at time $t$\cite{wu2020VariationalApproachLearning,scherer2019VariationalSelectionFeatures}. In practice, a finite set of timeseries of observables is used to obtain an estimation of the Koopman operator in a desirably small subspace relevant to slow kinetics. The eigenvalues and eigenvectors of such approximate Koopman operator can then be used 
to describe slow relaxation modes together with their corresponding characteristic timescales. The accuracy of the approximation can however strongly depend on the choice of observables.

In some cases, an intuitive reaction coordinate can be easily determined. For instance, two torsional angles of the backbone of the alanine dipeptide as the reaction coordinates are sufficient to describe the slow dynamics. In cases where intuitive reaction coordinates cannot be determined, many possible observables including root mean square displacement (RMSD), all possible torsional angles, native contacts, and backbone distances can be considered \cite{scherer2019VariationalSelectionFeatures}.
Depending on the target system's complexity, these generic observables may not form a good basis to extract transition timecales, requiring more elaborate schemes to define a better observable \cite{lindorff-larsen2011HowFastFoldingProteins}. Identifying a compact yet sufficiently-complete set of observables that is
able to reliably approximate the true relevant eigenfunctions of the Koopman operator remains an outstanding challenge, although guidelines are gradually emerging \cite{scherer2019VariationalSelectionFeatures, best2013NativeContactsDetermine}. 

To understand this challenge better, we can define the properties of an optimal set of observables. Mathematically, optimal observables should effectively act as basis functions to approximate the eigenfunctions corresponding to a ``slow'' subspace that describes rare transitions (e.g., protein folding/unfolding) or slowest relaxation modes of the exact Koopman operator \cite{wu2017VariationalKoopmanModelsb}, which are of course {\em a priori} unknown. While general eigenfunctions can be extremely complex, those that represent slow transition between $M$ metastable states have simplified features: they can be shown to be collectively approximated by linear combinations of $M$ indicator functions, each of which takes non-zero values over one of the metastable sets and zero otherwise \cite{PCCA2013}. Even when a good basis to extract such indicator functions is not available, it can be shown that these functions can be inferred by representing the evolution of non-ideal observables in terms of a Hidden Markov Model (HMM) with $M$ hidden states \cite{noe2013ProjectedHiddenMarkov}.

However, fitting an HMM model to high-dimensional data given an unknown $M$ number of hidden states is generally non-trivial, requiring iterative methods such as Expectation Maximization \cite{Baum1970,FrankHMM,Rabiner1989} or Sequential Monte Carlo methods\cite{smith2013sequential}, all of which are considerably more complex than the methods based on the traditional linear Koopman approaches. Note that the task of choosing ``right'' observables can be alleviated by using non-linear optimization methods such as neural networks \cite{VAMPnet,Yeung2017LearningDN} or kernel methods \cite{KernelTICA,Klus2019}. These methods often require very large amounts of data and careful regularization to avoid over-fitting, and hence require more expertise and careful application/fine-tuning than linear Koopman operator methods, which possess an appealing simplicity of use and interpretation.

In this study, we propose a simple and scalable alternative to traditional HMM-inference algorithms based on the two-sample Kolmogorov--Smirnov (KS)\cite{smirnov1948TableEstimatingGoodness,dimitrova2020ComputingKolmogorovSmirnovDistribution} test and agglomerative clustering \cite{sokal1962COMPARISONDENDROGRAMSOBJECTIVE}, hereinafter referred to as \emph{KS clustering}. This KS clustering is used to identify good ``surrogate'' observables that conceptually correspond to indicator functions over hidden/metastable states. The key idea is that the {\em statistics} of the time evolution of even imperfect observables should contain information that can be used to distinguish the metastable states a system visits, hence allowing one to infer a good basis for slow eigenfunctions from imperfect observables.


It is worth noting an important trade-off in this algorithm: while it produces accurate characteristic timescales, the resulting surrogate observables are not explicit functions of the degrees of freedom of the target dynamical system, and so the eigenfunctions that are produced cannot directly be interpreted mechanistically. The KS clustering however possess an advantage over methods where HMM membership functions are defined in the observable space, since it can implicitly construct surrogate functions that cannot be explicitly expressed in the observable space.


This article is organized as the follows. In Sections \ref{sec:Koopman} and \ref{sec:dataKoopman}, we provide theoretical background for the representation of the Koopman operator that can be estimated from time series of observables obtained from stochastic systems. Section \ref{sec:HMMconcepts} describes key concepts in HMM with multiple states that can be included in indicator functions. Section \ref{sec:TheoryClustering} explains further why indicator functions with the KS clustering algorithm work best in a reduced observable space for defining hidden states. Section \ref{sec:2HMMTheory} illustrates the computation of the Koopman operator with an HMM of two hidden states. In Section \ref{sec:NResults}, we demonstrate numerical results for the two-state HMM (Sec. \ref{sec:2HMMNumerical}) and apply the KS clustering to three protein systems (Sec. \ref{sec:Proteins}). Finally, we discuss some implications of the results in Section \ref{sec:DisCon}.

\section{Theoretical background}
\label{sec:Theory}

\subsection{Koopman representation of dynamical systems} \label{sec:Koopman}

Throughout this manuscript, we consider Markovian stochastic dynamics driving the evolution of a thermal system whose microscopic state is denoted as $\omega$ in a state space $\Omega$. For example, for a three-dimensional $N$-atom molecular system with $\Omega=\mathbb{R}^{3N}$ evolving under overdamped dynamics, the microscopic state can be fully characterized by $\omega=(x_1, \ldots, x_N, y_1\ldots, y_N, z_1, \ldots, z_{N})$, which are the Cartesian coordinates of all atoms. In the stochastic setting, an ensemble of trajectories at time $t$ is characterized by  $\rho(t,\omega)$ as a joint probability density function in the continuous state space $\Omega$. 

The dynamics is prescribed by an infinitesimal generator $\mathcal{L}$. To fix ideas, we consider overdamped Langevin dynamics with Gaussian white noise, $\mathcal{L}=- \sum_{i=1}^{3N} \l(\partial_{\omega_i} V\l(\omega\r)\r)\partial_{\omega_i} + 2k_B T \partial_{\omega_i}^2$, where $V(\omega)$ is the potential describing the interactions between the atoms.  For simplicity, we consider systems with detailed balance, which guarantees reversibility \cite{klus2018DataDrivenModelReductiona}. Note that the formalism itself is not specific to Langevin dynamics, but is applicable to general reversible dynamics. The infinitesimal generator $\mathcal{L}$ uniquely defines the evolution of the probability density function,
\begin{equation}
    \frac{\partial}{\partial t} \rho\l(t,\omega\r) = \mathcal{L}^\dagger \rho\l(t, \omega\r),
\end{equation}
where $\mathcal{L}^\dagger$ is the adjoint of $\mathcal{L}$. For reversible dynamics, $\mathcal{L}=-\mathcal{L}^\dagger$.
We assume that the stochastic system is ergodic such that a unique stationary distribution, $\rho_\text{stat}$, exists and satisfies $\mathcal{L}^\dagger \rho_\text{stat}(\omega)=0$. 

Let's consider an observable $O$ that is a real-valued function of $\omega$ and define the so-called stochastic Koopman operator $\mathcal{K}_t$ \cite{wu2020VariationalApproachLearning} as
\begin{equation}
\l(\mathcal{K}_t O\r)\l(\omega\r) \triangleq \mathbb{E}\l[O\l(\omega_{t}\r) \vert \omega_0 = \omega\r] , \forall \omega \in \Omega,
\label{eq:koopman_def}
\end{equation}
where $\omega_t$ and $\omega_0$ denote the stochastic processes measured at time $t$ and $0$, respectively. Using the semi-group notation, the finite-time Koopman operator $\mathcal{K}_t=e^{t\mathcal{L}}$ maps the current value of an observable $f$ to its expectation over the probability distribution induced by the process $O_t$ at a later time $t$, given the initial process $O_0$. A function $\phi$ is defined as a Koopman eigenfunction if it satisfies $\l(\mathcal{K}_t  \phi\r)= e^{\lambda t} \phi$ or equivalently  $\mathcal{L} \phi = \lambda \phi$ \cite{Koopman315,Koopman255,neumann1932ZurOperatorenmethodeKlassischen,mezic2005a}. 
In this study, we consider systems with point spectra, that is, systems with countable $\lambda_i$, $i=1\ldots $, which can be ordered by their moduli, whose corresponding eigenfunctions $\phi_i$ satisfy ~$\mathcal{L} \phi_i = \lambda_i \phi_i$.


\subsection{Data-driven estimation of the Koopman operator} \label{sec:dataKoopman}

The linearity of $\mathcal{K}_t$ and the correspondence between its eigenvalues/eigenfunctions and those of the generator $\mathcal{L}$ discussed  in Sec.\ \ref{sec:Koopman} can be leveraged to create powerful data-driven methods to efficiently learn the characteristics of dynamics from timeseries of observables.  Methods such as TICA \cite{molgedey1994SeparationMixtureIndependenta}, VAMP \cite{wu2017VariationalKoopmanModelsb,wu2020VariationalApproachLearning} and EDMD \cite{williams2015data} provide linear finite-dimensional  approximations to the Koopman operator acting in the space of selected observables. An estimated Koopman operator can be obtained by minimizing the $L^2$-norm of the difference between the left- and right-hand sides of Eq.~\eqref{eq:koopman_def} using a least-squares minimization over pairs of configurations separated by a lag time $t$.

For brevity, we consider discrete-time Markov processes below, noting that it is straightforward to generalize the analysis to continuous-time Markov processes. In the discrete-time setting, an approximation to the Koopman operator\cite{williams2015data,lin2021datadriven} given a sample path $\l\{\omega_t\r\}$, $t=0,1,\ldots$, and a vector-valued observable $O$ is given by:
\begin{equation}
   K_k \equiv C(k) \cdot C^{-1}(0),
   \label{eq:edmd}
\end{equation}
where $C(k)$ with $k \in \mathbb{N}$ is the $k$-lag correlation of the observable
\begin{equation}
    C(k) := \frac{1}{T} \sum_{s=0}^{T-1} O(\omega_{s+k}) \otimes O(\omega_s), 
\end{equation}
where the dyadic (outer) product is denoted by $\otimes$. 

The eigenvalues of $K_k$ are approximations of $e^{\lambda_i k}$, and its eigenvectors $\Phi_i$ can be used to approximate the eigenfunction $\phi_i$ of the true Koopman eigenfunctions defined in Sec.~\ref{sec:Koopman}\cite{williams2015data,lin2021datadriven} as $\phi_i(\omega_s) \simeq O(\omega_s) \cdot \Phi_i $, where $\cdot$ denotes the inner product of the two vectors\cite{williams2015data}. 
This simple and elegant procedure can be shown to converge to the exact eigenvalues and eigenfunctions in the limit of infinitely long timeseries of a set of observables $f$ which linearly span a Koopman invariant subspace containing the corresponding set of Koopman eigenfunctions $\l\{\phi_i\r\}_i$ \cite{Brunton2016Invariant,lin2021datadriven}. Based on the Rayleigh-Ritz method, Wu and No\'e \cite{wu2020VariationalApproachLearning} showed that the data-driven estimation of the Koopman operator is a variational problem when using methods such as TICA\cite{molgedey1994SeparationMixtureIndependenta,schwantes2013ImprovementsMarkovState,perez-hernandez2013IdentificationSlowMoleculara,naritomi2011SlowDynamicsProtein} and EDMD \cite{williams2015data}, with the approximate characteristic timescales approaching the actual values from below in the infinite-data limit.  When observables are chosen as the first $M$ Koopman eigenfunctions, $\l\{\phi_i\r\}_{i=1}^M$, the variational bound is tight,  thus recovering the optimal estimates of the Koopman eigenvalues and eigenfunctions. 
In practice, selecting a finite number of observables is the only option. This selection procedure is often system-specific, relying on educated guesses or {\em a priori} information of an intuitive reaction coordinate. The quality of the estimate can depend sensitively on a choice of observables, as we now show.

\subsection{Hidden Markov Chain as an effective model for describing systems with multiple metastable states} \label{sec:HMMconcepts}

In the following, we focus on the problem of characterizing the kinetics of systems with metastable states, which are defined to have distinguishable statistics,  essential to the approximation of the Koopman operator discussed in the previous sections.
Let's consider a system with $M$ such metastable states and denote the $i^\text{th}$ metastable state, $i=1\ldots M$, by $\Omega_i$ with $\Omega = \cup_{i=1}^M \Omega_i$, and $\Omega_i \cap \Omega_j=\varnothing$ if $i\ne j$.
Conceptually, a metastable state is such that a typical trajectory would relax to quasi-stationary distribution (QSD) within one such state much faster than it would leave the state \cite{le2012mathematical}. Many systems in biology, chemistry, and materials science exhibit strong metastability, which makes their study using direct simulation methods such as molecular dynamics challenging due to the long waiting times between state-to-state transitions.
This setting implies the existence of a slow subspace containing $M$ slow eigenvalues, well separated or statistically distinguishable from the rest of the spectrum. As discussed above,
indicator functions of the form:
\begin{equation}
    \mathbf{1}_{\Omega_i}(\omega):= \l\{ \begin{array}{ll} 
    1,& \text{if } \omega \in \Omega_i, \\
    0,& \text{else},
    \end{array}\r.
    \label{eq:indicators}
\end{equation}    
would form an excellent approximate basis for the {\em global} eigenfunctions of the slow subspace of the Koopman operator $\mathcal{K}$. 

Define the \emph{local} infinitesimal operator $\mathcal{L}_i^\dagger$ as the operator describing the dynamics on $\Omega_i$, applying absorbing boundary conditions on the boundary $\partial \Omega_i$ of $\Omega_i$, as in Ref.\ \cite{le2012mathematical}.
Then, the largest eigenvalue and eigenfunction pair satisfies
\begin{equation}
    \mathcal{L}_i^\dagger \rho_1^{(i)} \l(\omega\r) = \lambda_1^{(i)}  \rho_1^{(i)} \l(\omega\r),\quad i=1\ldots M,
\end{equation}
where $\lambda_1^{(i)}<0$ implies decaying dynamics and $\rho_1^{(i)}\l(\omega\r)=0$ if $\omega \notin \Omega_i$. The eigenfunction $\rho_1^{(i)}(\omega)$ is referred to as the quasi-stationary distribution (QSD) on state $\Omega_i$. Note that the $\lambda_1^{(i)}$ are now {\em local} quantities, in contrast to the discussion in the preceding sections that focus on the eigenvalues of the {\em global} generator. Specifically, the eigenvalue $\lambda_1^{(i)}$ quantifies the expected timescale $-1/\lambda_1^{(i)}$, over which a system residing in state $\Omega_i$ would finally escape\cite{le2012mathematical}. 
The metastability of the state can be quantified by the ratio of the expected escaping time to the relaxation time to the QSD, i.e., $(\lambda_2^{(i)}-\lambda_1^{(i)})/\lambda_1^{(i)}$. In the following, we assume that all states are sufficiently metastable so that $(\lambda_2^{(i)}-\lambda_1^{(i)})/\lambda_1^{(i)} \gg 1$, $\forall i$.

Let's suppose we periodically observe the a trajectory of the system  on some timescale 

\begin{equation}
 -1/(\lambda_2^{(i)}-\lambda_1^{(i)}) \ll  \tau  \ll -1/\lambda_1^{(i)}
\label{eq:qsd_sample_time}
\end{equation}
We can set the timescale $\tau=1$ by choosing an appropriate unit of time. With a probability $\sim 1-\exp(\lambda_1^{(i)}) \lesssim 1$, we would observe that the system remains in state $\Omega_i$; then, by construction, the next observed configurations would be  sampled from a distribution approaching the QSD on state $\Omega_i$. Alternatively, with a probability $\sim \exp(\lambda_1^{(i)}) \ll 1$, we would observe that the system escaped to another state $\Omega_j\ne \Omega_i$, where its configurations will be sampled from the QSD on state $\Omega_j$. 
Crucially, when observed on the timescale $\tau=1$ that is large compared to the internal relaxation time within basins, the time series of any observable should be well approximated by a sequence of i.i.d. random variables drawn from a distribution specific to the QSD of the metastable state where the system is currently trapped. In other words, such a time series will be well approximated by discrete-time Hidden Markov Model (HMM) with $M$ hidden states, where the approximation becomes increasingly good as each state becomes increasingly metastable and statistically distinguishable (see Ref. \cite{noe2013ProjectedHiddenMarkov} for an in-depth analysis). 
In this representation, transitions between hidden states are described by a discrete-time stochastic matrix $\pr{S_{t+1} = i\vert S_{t}=j}$ with $t=0,1,\ldots$ being the observation times and $S_t$ being a hidden state realized in random processes. In HMM, we cannot directly measure $S_t$, but rather, we measure an observable $O_t$ according to an observation model $\pr{O_t \vert S_t }$. In our case, the observation model depends on the QSD measured on each state, i.e., $O_t = O(\omega_t)$, where $O$ is a prescribed deterministic function of the system's state, and $\omega_t$ is distributed according to the QSD of the states in which $S_t$ resides.

This discussion highlights the important property of generic physical observables: if hidden metastable states exist as described above, the statistical distribution of generic observables should contain useful information that can be used to implicitly reconstruct indicator functions over hidden states. We now show how the statistics of these variations following transitions between states dictate how well the corresponding slow timescales can be estimated from the optimized Koopman operator.  

\subsection{Kolmogorov--Smirnov Clustering} \label{sec:TheoryClustering} 

\begin{figure}[!t]
    \centering
    \includegraphics[width=1\textwidth]{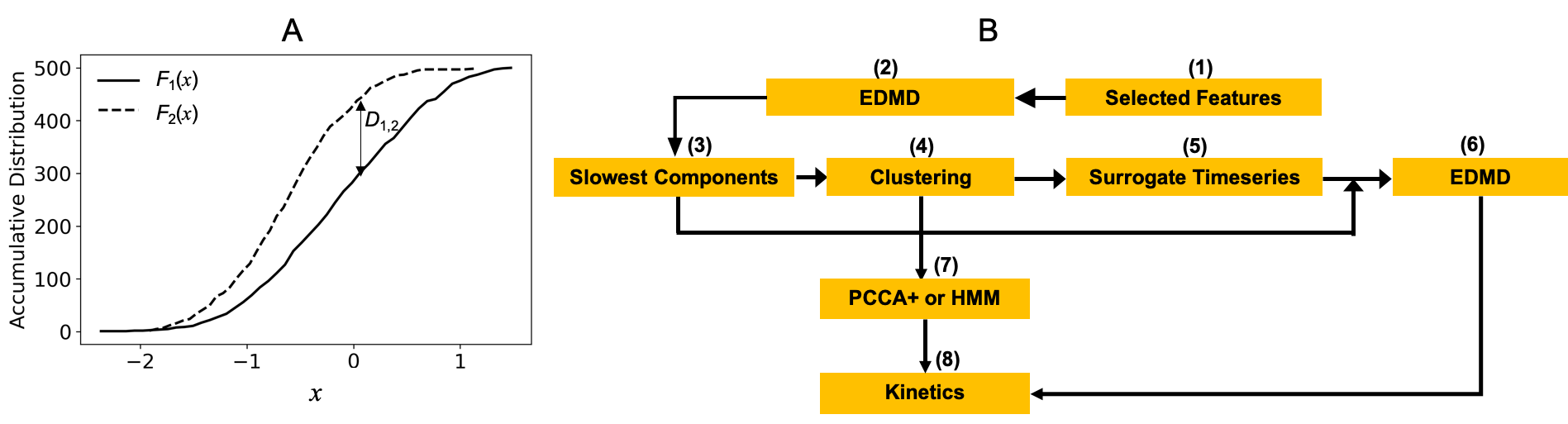}
    \caption{(Panel A) Illustration of the KS test. The test statistics corresponds to the maximum difference between two cumulative distribution functions measured. (Panel B) Schematic illustration of the proposed KS-clustering approach.}
    \label{fig:schematics3}
\end{figure}

The challenge in practice is to find a set of observables that can approximate the indicator functions over metastable states as characterized in Sec.~\ref{sec:HMMconcepts}. 
Our approach to this challenge is based on a simple characteristics of HMM: if during two time intervals $[t_1,t_1+\Delta t]$ and $[t_2,t_2+\Delta t]$ the dynamical system is in the same metastable state and delays between observations obey Eq.~\eqref{eq:qsd_sample_time}, then the distributions of observables measured over these time intervals should be statistically equivalent; and if the system is in different metastable states during the two intervals, the corresponding distributions of observables can presumably be statistically distinguishable from one another. The ``statistical distance'' between distributions measured in different time intervals can be quantified using the two-sample Kolmogorov--Smirnov (KS) test \cite{smirnov1948TableEstimatingGoodness}. Namely, $D_{1,2} = \sup_{x } |F_1(x)-F_2(x)|$ (Figure \ref{fig:schematics3}a) is the KS statistic measuring the maximum difference between two empirical cumulative distributions $F_1(x)$ and $F_2(x)$, which are computed from the data collected in two time intervals $[t_1,t_1+\Delta t]$ and $[t_2,t_2+\Delta t]$, respectively. A so-called distance matrix contains all $D_{i,j}$ values, which are computed for all pairs of intervals. Based on this distance matrix, indicator functions can be constructed (see below). 


In conventional applications, the KS statistics are used to reject the null hypothesis that the two samples were drawn from the same underlying distribution. In the present context, the KS statistics of the distributions corresponding to every pair of intervals are instead used as a statistical distance measure that allows for the clustering of all intervals into a number of different groups. Specifically, each group will contain intervals that are statistically similar to one another, while intervals that are very statistically different will be assigned to different groups. This can be done with any clustering methods that can operate from a user-provided pairwise distance matrix; in the following, this was accomplished via hierarchical agglomerative clustering\cite{bouguettaya2015EfficientAgglomerativeHierarchical}, which returns a hierarchy of clusters, i.e., clusters being merged in a bottom-up fashion until a preset number of cluster or a critical inter-cluster distance threshold has been reached.

Assuming that $M$ different groups are identified by such clustering, we can build $M$ surrogate timeseries conceptually corresponding to indicator functions over metastable states (rigorously, only $M-1$ such timeseries are needed, since the $M^\mathrm{th}$ one can be expressed as a linear combination of the $M-1$ others, up to an additive constant). Surrogate indicator function $l_i$ with $1\leq i \le M$ can be created by assigning a value of $1$ to a given time interval when it was deemed a member of cluster $i$ and a value of zero otherwise. This procedure can easily be generalized to multiple observables using multi-dimensional generalizations of the KS test \cite{multiks}, or by defining the distance between two multi-dimensional distributions as the maximal distance between any corresponding pair of one-dimensional distributions, which we use in the following. 

The overall algorithm is illustrated in Fig.\ \ref{fig:schematics3}.
One first identifies a set of base observables (1) which are processed using a conventional linear EDMD procedure (2).
The original timeseries are then compressed into a lower-dimensional space via projection into the eigenvectors corresponding to the slowest relaxation modes (3). 
 In this reduced space, the KS clustering is applied to identify indicator functions over the KS clusters (4), which are then used to construct surrogate time-series (5).
 The projected descriptors from step (3) are combined with the surrogate functions from step (5) and used as input for a final EDMD analysis (6) to compute slow modes and corresponding timescales (7).

\subsection{A two-state HMM with Gaussian observation noise} \label{sec:2HMMTheory}

To illustrate the arguments presented in Sec.\ \ref{sec:Koopman}-\ref{sec:TheoryClustering}, we consider an analytically solvable HMM with two discrete state $S \in \l\{1,2\r\}$ and a single observable $O \in \mathbb{R}$ (note that this analysis can be generalized to general $M$-state HMM). We use the standard notation that the upper-case symbols with a subscript time stand for random processes, and lower-case symbols stand for dummy variables or sample paths of random processes. The Markov transition between the hidden states is characterized by a Markov matrix $\bf{M}$ whose entries $M_{ij}=\pr{S_{t+1}=i \vert S_{t}=j}$:
\begin{equation}
    \bf{M}:= \begin{bmatrix}
    1- p_+ & p_- \\
    p_+ & 1 - p_- 
    \end{bmatrix}. \label{eq:HMM1}
\end{equation}
That is, with a probability $p_+$ (resp.~$p_-$) the hidden state jumps from 1 to 2 (resp.~2 to 1) in a single step. We note that the stationary distribution of the hidden state $\pi := \l[\pi_1, \pi_2\r]^T = \l[p_-/\l(p_- + p_+\r), p_+/\l(p_- + p_+\r) \r]^T$ satisfies $\mathbf{M}\cdot \pi= \pi$. As we are interested in systems whose hidden states are metastable on the observation timescale, we have $p_+$, $p_-\ll 1$. We consider a univariate Gaussian observation model, where the observation $O_t$ at time $t$, which is a random variable (modeling the quasi-stationary distribution), depends on only on the current hidden state $S_t$: 
\begin{align}
    \rho\l(O_t=\omega \vert S_t = s\r)  ={}& \frac{1}{\sqrt{2\pi \sigma_{s}^2 }} e^{-\frac{\l(\omega-\mu_{s}\r)^2 }{2 \sigma_{s}^2 }} \label{eq:HMM2}
\end{align}
where $(\mu_1, \sigma_1)$ and $(\mu_2, \sigma_2)$  fully characterize the observation model. We remark that the only timescale of the process is the autocorrelation time of the hidden states, which is $-1/\log(1-p_- - p_+)$. Without loss of generality, we impose a zero-mean condition on the observable, that is, $\lim_{T\rightarrow \infty} \frac{1}{T} \sum_{i=0}^{T-1} \omega_i =  \pi_1 \mu_1 + \pi_2 \mu_2 = 0$. 

Our goal is to analytically express the result of an EDMD procedure given an infinitely long timeseries $O_t$'s. This corresponds to substituting the analytical expression
\begin{equation}
    C(k) := \int_{-\infty}^{\infty } \int_{-\infty}^{\infty} \sum_{s,s'\in\l\{1,2\r\}}\, \pr{\omega_2 \vert s'} \pr{\omega_1 \vert s}\, \l[\mathbf{M}^k \r]_{s',s}\, \pi\l(s\r)\, \dd \omega_1 \, \dd \omega_2.  
\end{equation}
into Eq.\ \ref{eq:edmd}. It is elementary to show that 
\begin{equation}
    C(0) =  \pi_1 \l(\mu_1^2 + \sigma_1^2 \r) + \pi_2 \l(\mu_2^2 + \sigma_2^2 \r)
\end{equation}
and for $k \in \mathbb{Z}_{\ge 1}$, 
\begin{equation}
    C(k) = \l(1-p_+-p_-\r)^k \l(\mu_1 - \mu_2 \r)^2 \pi_1 \pi_2.
\end{equation}
Consequently, the estimated Koopman operator using this single observable with $k$-lag is 
\begin{equation}
    K_k = \gamma \l(1-p_+-p_-\r)^k, \label{eq:Koopman1}
\end{equation}
where 
\begin{equation}
    \gamma := \frac{ \l(\mu_1 - \mu_2 \r)^2 \pi_1 \pi_2 }{ \pi_1 \l(\mu_1^2 + \sigma_1^2 \r) + \pi_2 \l(\mu_2^2 + \sigma_2^2\r)}. \label{eq:gamma}
\end{equation}
The only characteristic timescale of the process is then estimated as
\begin{equation}
    \tau_k = - \frac{k}{\log K_k}  = \frac{1}{-\log \l(1-p_+-p_-\r) + \frac{1}{k} \log \gamma }. \label{eq:tauk}
\end{equation}


For general $\sigma_1$, $\sigma_2 \ne 0$ and a finite lag $k<\infty$, $\gamma <1$ leads to an overestimation of the timescale $\tau_k$. 
This result is consistent with the variational principle \cite{wu2020VariationalApproachLearning} which states that sub-optimal observables result into overestimations of the characteristic timescales; 
$-\log(\gamma)$ can be seen as a noise-to-signal metric which discounts the timescale estimation by EDMD. While in principle the correct timescale is recovered in the limit $k\rightarrow \infty$, approaching this limit could require extremely long trajectories.
In contrast, using zero-mean timeseries of the indicator function 
over the hidden states as a basis results in an unbiased estimation of $\tau_k=-1/\log\l(1-p_+-p_-\r)$ for any value of $k$.

\section{Numerical Results} \label{sec:NResults}
\begin{figure}[!t]
    \centering
    \includegraphics[width=1.0\textwidth]{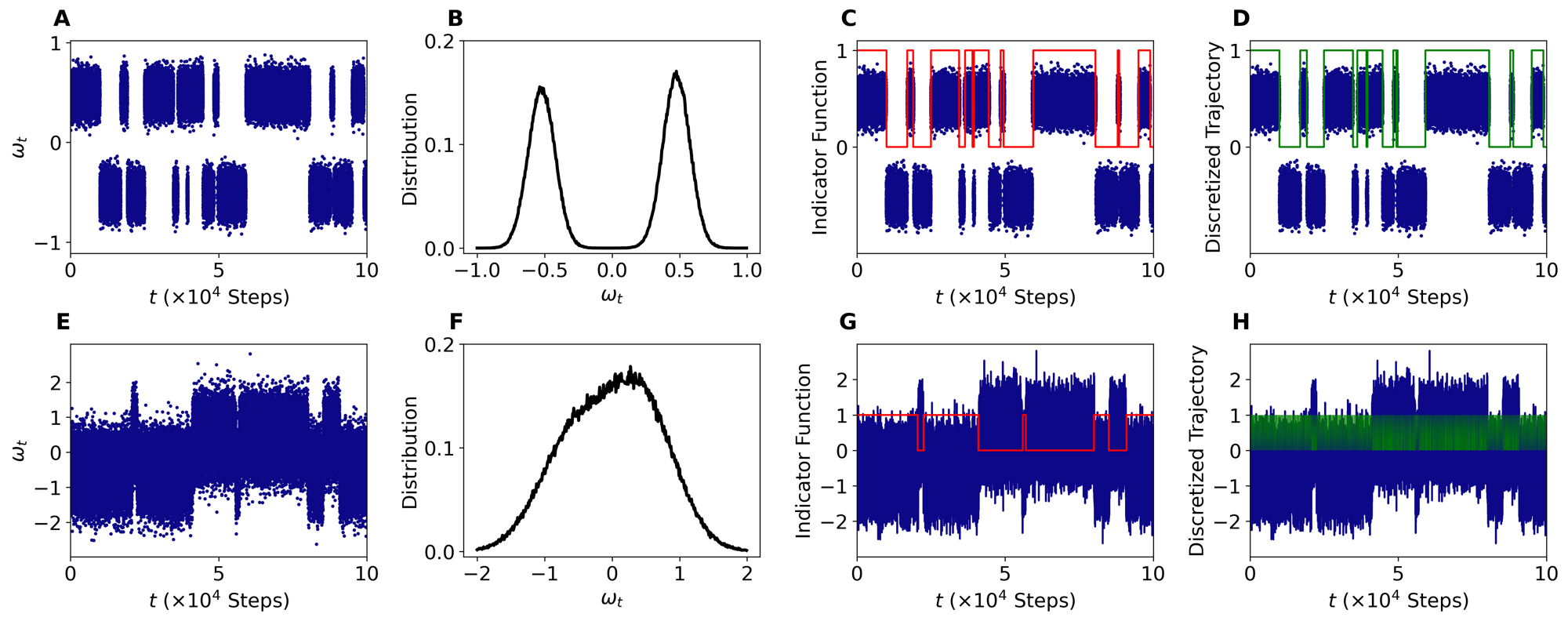}
    \caption{Illustration of the KS-clustering approach using timeseries generated with the Hidden Markov Models defined in Eqs.~\eqref{eq:HMM1} and \eqref{eq:HMM2} with $p_+=p_-=10^{-4}$ and $\mu_1=-0.5$, $\mu_2=0.5$. Algorithmic parameters $\Delta t=500$ for the KS clustering procedure. Panels A-D: $\sigma_1=\sigma_2=0.1$; Panels E-H: $\sigma_1=\sigma_2=0.5$. Panels A and E are the time series generated by a standard kinetic Monte Carlo sampling procedure, and the histogram of the time series are shown in panels B and F. The red-colored indicator function based on the KS clustering is illustrated in Panels C and G, as well as green-colored indicator functions (often called discretized trajectories) based on a standard $k$-means clustering in Panels D and H. \cite{pande2010EverythingYouWanted,Pande2013,scherer2015PyEMMASoftwarePackage}} 
    \label{fig:HMM}

\end{figure}
\subsection{Two-state HMM with Gausssian noise} \label{sec:2HMMNumerical}

We now use the two-state HMM discussed above to numerically demonstrate that the approximated indicator functions inferred from the KS clustering significantly improve the timescale estimation. For each set of parameters considered, a reference trajectory of $5\times 10^5$ steps is first generated using a standard kinetic Monte Carlo procedure. In the following, the transition probabilities between the hidden states was fixed at $p_+=p_-=10^{-4}$; and the conditional means of the observation were set to $\mu_1=-0.5$ and $\mu_2=0.5$, ensuring that the long-time observable mean tends to 0. In the first parameter set, we consider $\sigma_1=\sigma_2=0.1$, corresponding to a well-separated observation model (see time series in Fig.~\ref{fig:HMM}A and empirical distribution in Fig.~\ref{fig:HMM}B). In the second parameter set, we consider $\sigma_1=\sigma_2=0.5$; in this case the stationary distribution is unimodal (see time series in Fig.~\ref{fig:HMM}E and empirical distribution in Fig.~\ref{fig:HMM}F); this is representative of a situation where the base observables are poor at distinguishing the two states (which is a very common occurrence in practice). 

\begin{figure}[!t]
    \centering
    \includegraphics[width=1\textwidth]{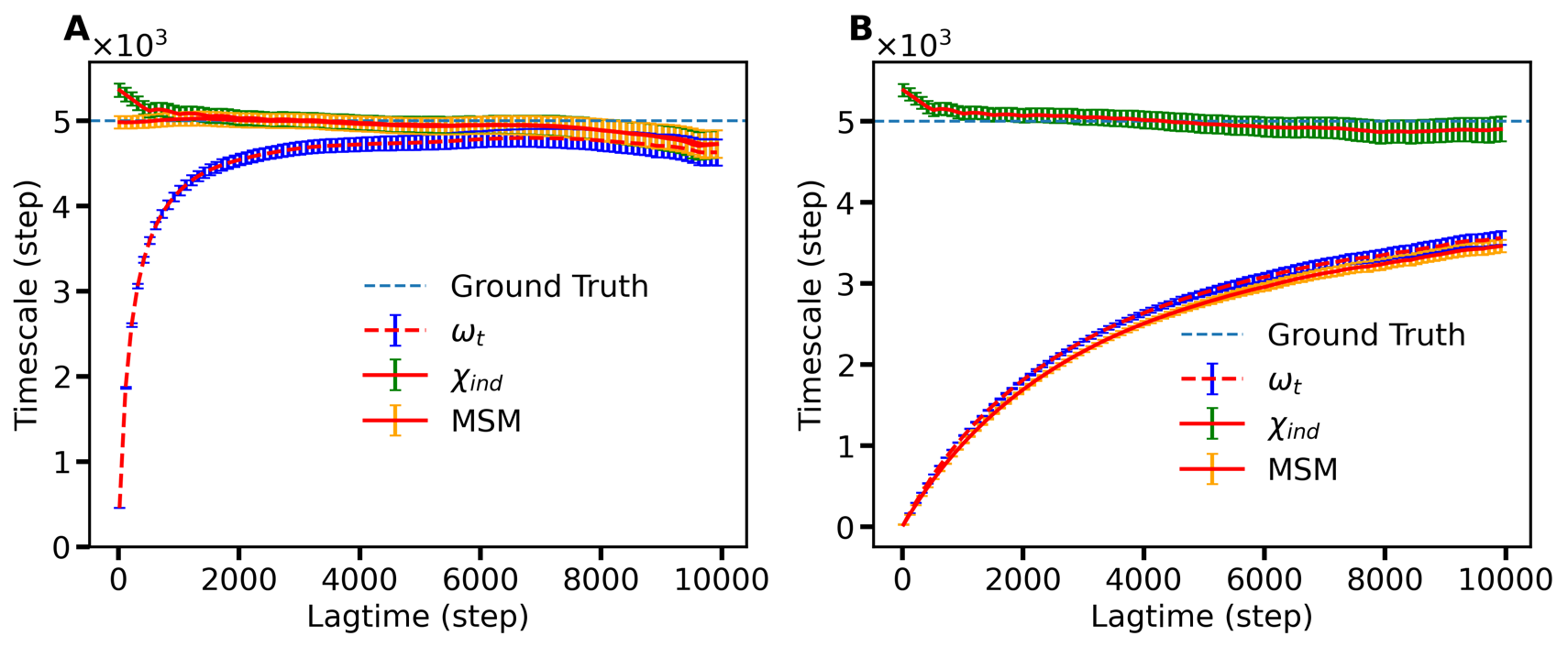}
    \caption{Timescales (Left axis) 
    computed for the two-state Hidden Markov Models. The dashed and solid black or red curves are the data obtained via EDMD from  $\omega_t$ and $\chi_\text{ind}$, respectively. Panel A: $\sigma_1=\sigma_2=0.1$; Panel B: $\sigma_1=\sigma_2=0.5$. Label $\omega_t$ indicates the estimated timescales obtained from the raw data via EDMD without clustering. Label $\chi_\text{ind}$ indicates the timescales obtained from the indicator functions computed via the KS clustering applied to $\omega_t$. An ensemble of 100 time-series was used. Each timeseries has $5\times 10^5$ steps.}
    \label{fig:HMMVAMP}
\end{figure}

In both parameter sets, the ground-truth characteristic timescale is $-\log\l(1-p_1 - p_2\r) \approx 5000$ discrete time steps. The discounting 
factors $\gamma$ from Eq.~\eqref{eq:gamma} are $1/1.04$ and $1/2$, respectively. These two cases are thus representative of good and bad descriptors, where the convergence of EDMD with respect to the lag time would be fast and slow as shown by the timescales (dashed red curve) in Fig.\ \ref{fig:HMMVAMP}A and B, respectively. Even for the ``good'' case (Fig.\ \ref{fig:HMMVAMP}A), the slowest characteristic timescale obtained from $\omega_t$ (raw data used for step (2) in Fig.\ \ref{fig:schematics3}B) is severely underestimated for lag times less than 1000 steps. The estimate is even worse in the ``bad'' case (Fig.\ \ref{fig:HMMVAMP}B), where a 20\% underestimation still persists even for a lag of 10,000 steps.

We now demonstrate that the use of the KS indicator functions can greatly improve the estimated timescales. To obtain the KS indicator functions  (see Sec.\ \ref{sec:TheoryClustering}), we applied the procedure (steps 1-6) described in Fig. \ref{fig:schematics3}b. At step (4), the data were split into time intervals of $\Delta t=500$, from which the inter-interval KS distance matrix was computed. This matrix was then used as an input to the hierarchical  agglomerative clustering algorithm \cite{bouguettaya2015EfficientAgglomerativeHierarchical} with $M=2$ clusters. The KS indicator functions are plotted in Fig. \ref{fig:HMM}C and G for both cases. 
Note that since the dimension of the test system is one, steps (1-3) can be skipped. As a point of comparison, we also considered a standard MSM approach using  k-mean clustering with 2 clusters; the corresponding discretized trajectories are illustrated in Fig. \ref{fig:HMM}D and H. For the ``good'' case, Figures \ref{fig:HMM}C and D show consistent indicator functions obtained from both clustering algorithms. 
Fig.\ \ref{fig:HMMVAMP}A shows that both of the clustering algorithms yield  accurate timescales at all lagtimes.
When the distributions corresponding to the two hidden states are not well separated (c.f. Fig. \ref{fig:HMM}E and F), Fig. \ref{fig:HMMVAMP}B shows that the accuracy of the estimated timescale obtained from the standard MSM with k-means clustering is basically identical to the direct EDMD approach, while the KS-clustering approach accurately estimates the slow timescale at all lagtimes. This results from the fact that the KS clustering uses observable statistics over extended periods of time to differentiate metastable states, while a k-means approach operates on each sample separately, and hence cannot distinguish distributions that significantly overlap with one another.

\begin{figure}[!t]
    \centering
    \includegraphics[width=0.96\textwidth]{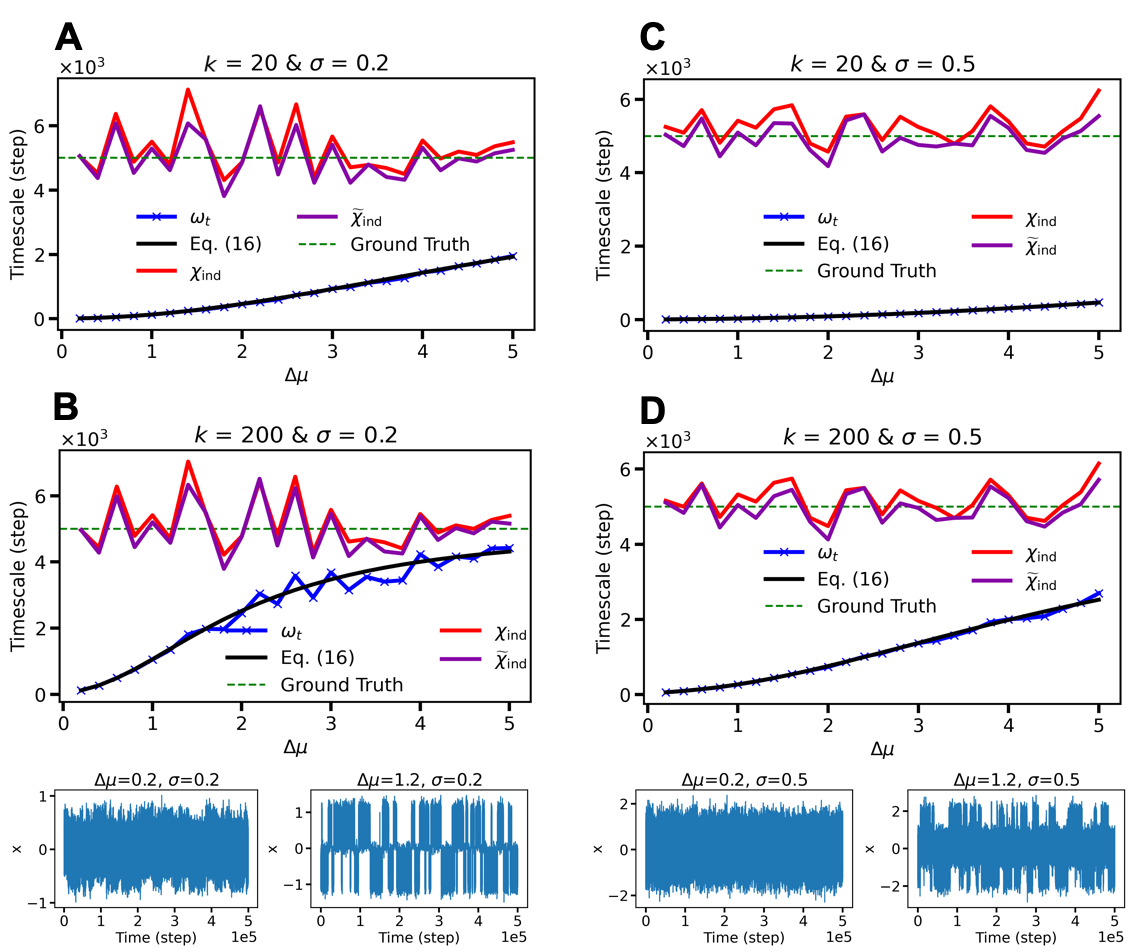}
    \caption{Timescale ($\tau_k$) computed for different $\sigma_1=\sigma_2=\sigma$ and lagtime $k$ as functions of $\Delta \mu=\mu_2-\mu_1=2\mu_2$, fixing $p_1=p_2=10^{-4}$. The dashed green line is the ground true timescale, $\tau_\text{truth}=5000$ steps. Here, we used the exact indicator functions ($\tilde{\chi}_{ind}$), which were generated by the kinetic Monte Carlo procedure, for comparison with the indicator function $\chi_{ind}$ generated by the KS clustering with number $M=2$ of clusters used for all calculations here.}
    \label{fig:varyingMu}
\end{figure}

\begin{figure}[!t]
    \centering
    \includegraphics[width=0.8\textwidth]{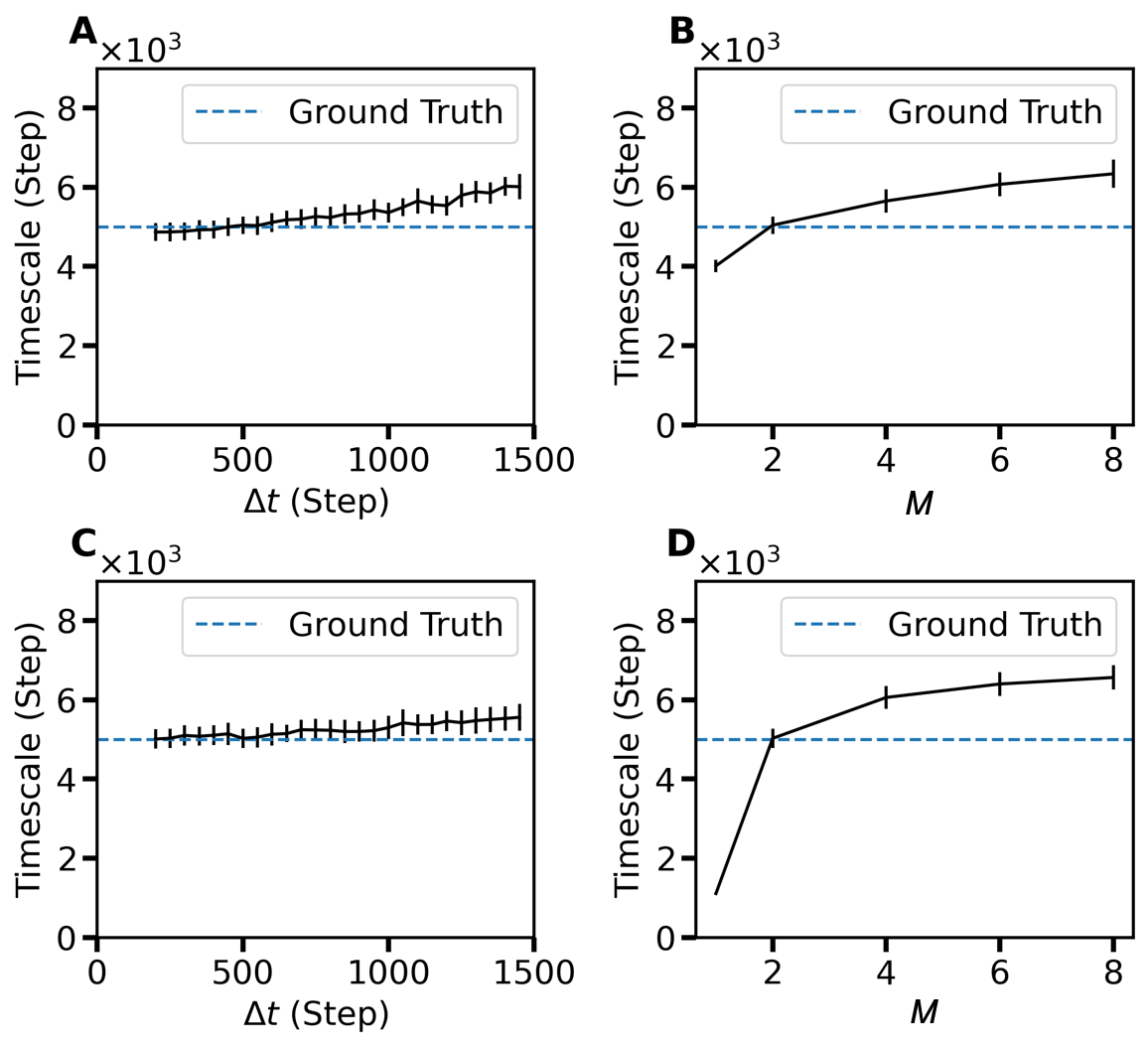}
    \caption{Dependence of timescales on $\Delta t$ and number of clusters ($M$), which is equal to the number of indicator functions, $\chi_\mathrm{ind}$. Panels A-B: $\sigma_1=\sigma_2=0.1$; Panels C-D: $\sigma_1=\sigma_2=0.5$; Panels A and C: $M = 2$; Panels B and D: $\Delta t = 500$.  Other parameters are the same as in Figure \ref{fig:HMM}. An ensemble of 10 timeseries was used to compute the confidence intervals. Each timeseries has $5\times 10^5$ steps.}
    \label{fig:dKS}
\end{figure} 

To further examine the effects of $\mu_i$ and $\sigma_i$ on the estimation of timescales, we fixed $p_+=p_-=10^{-4}$ and varied $\Delta \mu=\mu_2-\mu_1=2\mu_2$ and $\sigma_1=\sigma_2=\sigma$. 
Equation \eqref{eq:gamma}
then simplifies to $\gamma=1/\l[4(\sigma/\Delta\mu)^2+1\r]$. 
Figure \ref{fig:varyingMu} shows that the results from direct EDMD on $\omega_t$ are underestimated in a way that is in good agreement with the predictions of Eq.~\eqref{eq:tauk}. In contrast, the use of the indicator functions obtained from the KS clustering recovers reasonable timescale even at short lagtimes and when the noise amplitude is very large compared to $\Delta \mu$ in comparison with the exact indicator functions. The very good agreement between the results obtained from inferred indicator functions and actual hidden states (which are not normally accessible) indicates that 
the fluctuations in predicted timescales result from the finite trajectory lengths.

The KS clustering method requires selecting two tunable parameters: the length of the bins $\Delta t$ used to partition the timeseries and the number of target clusters $M$ used in the clustering algorithm. The impact of these two parameters is shown in Fig. \ref{fig:dKS}. Figures \ref{fig:dKS}A and C show that $\Delta t$ around 500 steps produces accurate results, while larger values lead to an overestimation of the characteristic timescale. This behavior can be rationalized by considering the $\Delta t >> \tau $ limit. In this case, multiple visits to different metastable states are averaged out within each interval. This results into an underestimation of the transition probability and a corresponding overestimation of the transition timescales. In order to be accurate, on one hand, the algorithm requires that most time intervals contain sections of trajectory remaining in the same hidden states, whose timescales satisfy Eq. \ref{eq:qsd_sample_time}. On the other hand, $\Delta t$ should not be too small (e.g., $< 50$ in our cases), so as to maximize the statistical power of the KS test for accurately detecting transitions between metastable states. We therefore recommend $\Delta t \simeq \tau_M/10 $, where $\tau_M$ is the timescale of the fastest mode in the target ``slow'' subspace. Of course, $\tau_M$ is not known {\em a priori}, so $\Delta t$ can be estimated using the EDMD preprocessing step discussed above. This estimate can be validated and readjusted after a carrying out the whole procedure.

Figures \ref{fig:dKS}B and D shows the effect of the number of clusters/hidden states $M$. 
These results suggest that the estimated transition timescales can slightly be overestimated when $M$ is larger than the number of hidden states (2 in this case), although the dependence is rather weak. Consider the extreme limit where 
 $M$ is equal to the total number of time intervals, and hence each interval will be assigned to its own cluster. In this case, it is possible for EDMD to create a spurious linear combination of indicator functions whose characteristic timescale approaches the trajectory length. We again recommend using the EDMD preprocessing step discussed above to estimate the number of slow characteristic timescales prior to the KS-clustering procedure.

\subsection{Application to Protein Dynamics} \label{sec:Proteins}

\begin{figure}[!t]
    \centering
    \includegraphics[width=1\textwidth]{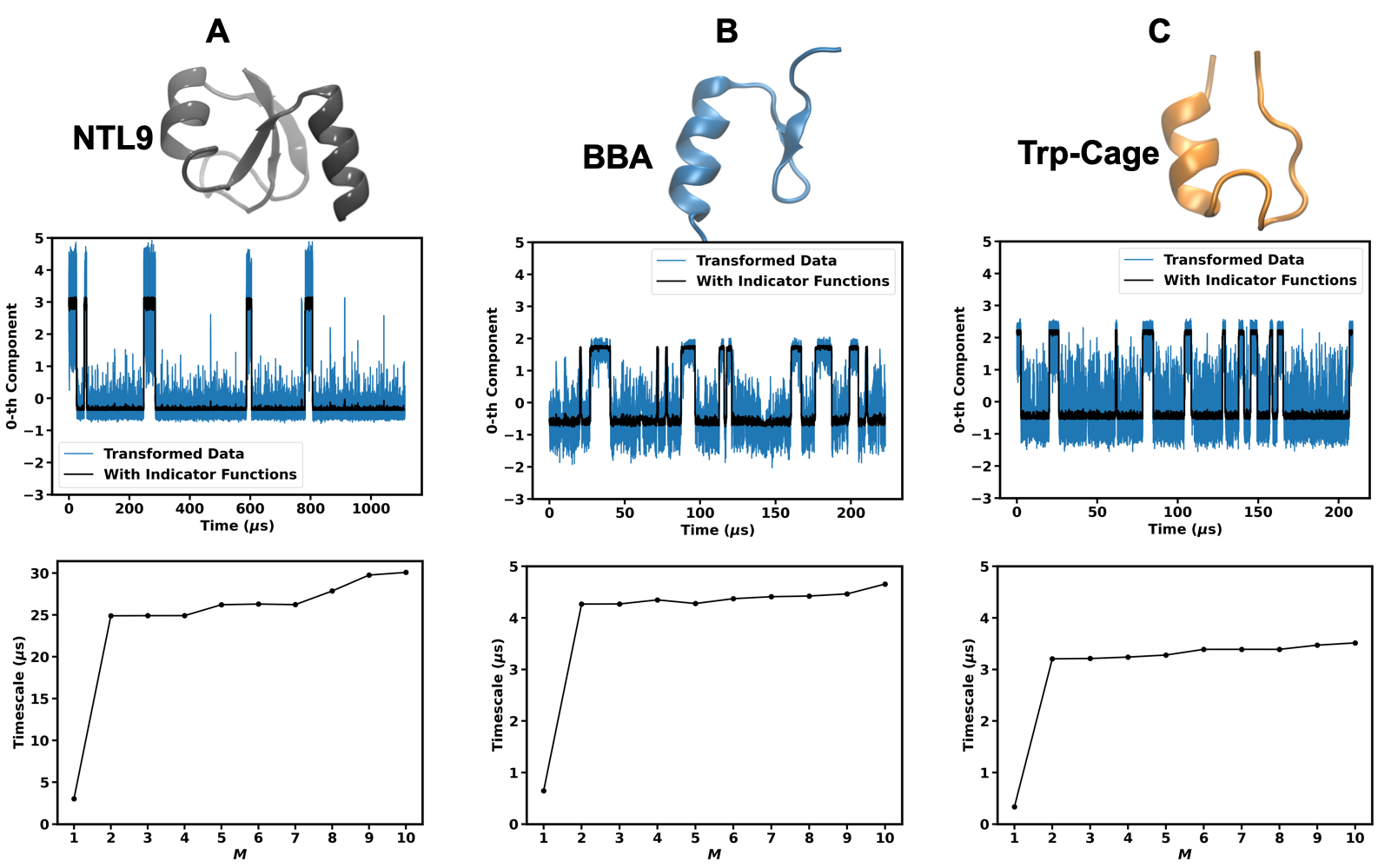}
    \caption{Protein systems. A: Ribosomal Protein L9 (NTL9) (PDB: 2HBB), B: Beta-beta-alpha Fold (BBA) (PDB: 1FME), and C: Trp-Cage (PDB: 2JOF). Lagtime = 100 steps or 200, 40 and 40 ns for NTL9, BBA and Trp-Cage, respectively;  $\Delta t = 2100 \text{ steps}$ or $4.2, 0.84, \text{and } 0.84\ \mu s$ for NTL9, BBA and Trp-Cage, respectively. Middle Panels: The timeseries of the first slowest components in the protein systems. Bottom Panels: the longest time scale as a function of selected components.}
    \label{fig:protein}
\end{figure}

To illustrate the performance of the KS clustering on more complex systems, we consider three small proteins, NLT9, BBA, and Trp-Cage (Fig. \ref{fig:protein}A-C), each of which was simulated for at least \SI{200}{\micro\second}\cite{lindorff-larsen2011HowFastFoldingProteins}.  We describe each trajectory by the time series of the torsional angles 
of each amino acid in the proteins, indexed with $i$ \cite{mcgibbon2015MDTrajModernOpen}, yielding  190, 152, and 85 observables, respectively. We then applied EDMD to estimate the Koopman operator for the three systems. The trajectories were then projected onto the $M$ slowest eigenvectors of the Koopman operator. The resulting $M$ timeseries were used as input features to the procedure (steps 1-6) shown in Figure \ref{fig:schematics3}B. Note that when $M=1$, no surrogate indicator function is introduced and the results are identical to that of a conventional EDMD approach.

Figure \ref{fig:protein} shows the projection of the original trajectory into the slowest eigenmode (middle row) of the Koopman operator inferred from the original data (blue) and from the augmented data. While the data remains relatively noisy when projected based on the torsional angles only (blue), the almost piecewise character becomes evident when EDMD is augmented by the surrogate indicator functions (black), which is a clear indication that improved eigenfunctions are produced (see Sec.~\ref{sec:2HMMTheory}). This strongly suggests that the KS clustering properly identifies the metastable states of the system. This improved identification dramatically affects the estimated timescales compared to the original EDMD (which corresponds to $M=1$), sometimes by up to an order of magnitude (c.f.~bottom row of Figures \ref{fig:protein}). The slowest characteristic timescale is also observed to be relatively insensitive to a choice of $M$, beyond a very large initial jump as soon as at least one surrogate indicator function is added (i.e., when $M\ge2)$. 

As a more stringent test of the approach, we repeated the same procedure using a single base observable, the RMSD with respect to the native state. Using only the RMSD in an EDMD approach produces poor estimates of the timescales, namely 0.25, 0.05, and \SI{0.1}{\micro\second} for NLT9, BBA, and Trp-Cage, respectively. 
When combined with indicator functions ($M=2$, see SI) for NLT9, BBA, and Trp-Cage, the timescales are increased to 22.6, 2, and \SI{3}{\micro\second}, respectively. These estimated timescales from the RMSDs are close to the estimates from a much richer set of observables \cite{lindorff-larsen2011HowFastFoldingProteins} and timescales shown in Fig. \ref{fig:protein}. This suggests that  the KS-clustering is comparatively much less sensitive to the details of the observable definition compared to the conventional EDMD approach, insofar the observable distributions in different states is statistically distinguishable.

\section{Conclusion} \label{sec:DisCon}

EDMD-type methods are powerful kinetic analysis tools that are particularly appealing due to their formal and practical simplicity. 
The results of this type of analysis are however often sensitive to the ability of a given set of observables to serve as a good basis for the slow subspace of the generator. This would even be true of non-linear generalization of EDMD or even conventional MSM approaches when the observable distributions corresponding to different hidden states are not separable in the space spanned by the observable.  Rationally enriching the observable space is an outstanding challenge that has yet to find a simple solution. 

The main objective of this work was to propose a strategy to retain the simplicity of linear EDMD-type methods, while improving the ability of the method to extract accurate kinetic information from relatively poor observables. The KS-clustering approach proposed here does not require the distributions to be separable in the observable space, but only to be sufficiently statistically distinct that those corresponding to different hidden states can be identified using a two sample KS test. The surrogate indicator functions obtained by clustering using the KS metric were shown provide reliable estimates of known characteristic timescales, even when the distribution of observables over the different metastable states strongly overlap in the given observable space. It is however worth nothing that the approximate eigenfunctions obtained by KS-clustering are not explicit functions of the observables, which is certainly a limitation in terms of their direct interpretability. However, the availability of timeseries ``labeled'' in terms of putative hidden states can potentially be correlated with known observables to interpret the nature of the transitions between hidden states.
When enriched with surrogate indicator functions over the implicit metastable set, the simple linear EDMD procedure is shown to produce very accurate timescale estimations at any lagtime, in contrast to conventional linear approaches that can require very long lagtimes to produce accurate estimates. The procedure is generic and scalable, and provides a simple tool to improve linear approaches at low cost.

\begin{acknowledgement}

This work has been authored by employees of Triad National Security, LLC which operates Los Alamos National Laboratory (LANL) under Contract No. 89233218CNA000001 with the U.S. Department of Energy/National Nuclear Security Administration. 
The work has been supported by LDRD (Laboratory Directed Research and Development) program at LANL under project 20190034ER (Massively-Parallel Acceleration of the Dynamics of Complex Systems: a Data-Driven Approach). V.A.N was partially supported by Director's Postdoctoral Fellowship, 20170692PRD4, for this work. V.A.N.~is supported by Oak Ridge National Laboratory, which is managed by UT-Battelle under Contract No. DE-AC05-00OR22725 with the U.S. Department of Energy. This research used resources of the Oak Ridge Leadership Computing Facility (OLCF). We also thank DE Shaw for making their MD data available
for this study.

\end{acknowledgement}

\begin{suppinfo}
Supporting information contains a PDF with one figure, and python scripts and pandas pickles, which are made available at https://gitlab.com/ngoav/the-ks-clustering.

\end{suppinfo}

\bibliography{achemso-demo}

\end{document}